\documentclass[aps,preprint,preprintnumbers,nofootinbib,showpacs]{revtex4}
\usepackage{graphicx,axodraw}
\begin{document}
\title{$Z$ decay into a bottom quark, a light sbottom and a light gluino}
\author{Kingman Cheung}
\email[Email:]{cheung@phys.cts.nthu.edu.tw}
\affiliation{National Center for Theoretical Sciences, National Tsing Hua 
University, Hsinchu, Taiwan, R.O.C.}
\author{Wai-Yee Keung}
\email[Email:]{keung@uic.edu}
\affiliation{Physics Department, University of Illinois at Chicago, 
IL 60607-7059 \\
National Center for Theoretical Sciences, National Tsing Hua 
University, Hsinchu, Taiwan, R.O.C.}
\date{\today}

\begin{abstract}
The discrepancy between the measured and theoretical production cross
section of $b$ quarks at the Tevatron can probably be explained by the
recently proposed scenario of light gluinos of mass $12-16$ GeV and
light sbottoms of mass $2-5.5$ GeV.  In this scenario, we study a
related process at the $Z$ pole, $Z \to b \tilde{b}_1^* \tilde{g} \, +
\, \bar b \tilde{b}_1 \tilde{g}$ followed by $\tilde{g} \to b
\tilde{b}_1^*\;/\; \bar b \tilde{b}_1$.  The hadronic branching ratio
for this channel is $(1-3) \times 10^{-3}$, which is of order of the
size of the uncertainty in $R_b$.  We find that a typical event
consists of an energetic prompt bottom-jet back-to-back with a ``fat''
bottom-jet, which consists of a bottom quark and two bottom squarks.
Such events with a $10^{-3}$ branching ratio may affect the
measurement of $R_b$; even more interesting if the ``fat'' bottom jet
can be identified.
\end{abstract}
\preprint{hep-ph/0207219}
\preprint{NSC-NCTS-020718}
\maketitle

\section{Introduction}

There has been a persisting discrepancy that 
the measured cross section of hadronic production of $b$ quarks 
measured by both CDF and D\O\ collaborations
\cite{cdf-d0} is about a factor of two larger than the prediction
in perturbative QCD with the most optimal choice of parameters,
such as $b$-quark mass ($m_b$) and the factorization scale $\mu$,
tuned to maximize the calculated rate.
\footnote{
Refs. \cite{bernd,cacc} argued that if the most up-to-date $B$ 
fragmentation function is used the observed excess can be reduced to
an acceptable level.  Fields \cite{field} interestingly pointed out
that correlations between the $b$ and $\bar b$ can be used to
isolate various sources of production, especially, in his study he 
included the fragmentation of gluon and light quarks.
}
Recently, Berger et al. \cite{berger} interpreted the discrepancy
in the scenario of light gluinos and light sbottoms.
Light gluinos of mass between $12-16$ GeV are pair-produced by QCD $q\bar q$ 
and $gg$ fusion processes, followed by 
subsequent decays of gluinos, $\tilde{g} \to b \tilde{b}^*_1 \,/ \, \bar b
\tilde{b}_1$, where the sbottom has a mass $2-5.5$ GeV. 
Therefore,
in the final state there are $b \bar b + \tilde{b}_1\tilde{b}^*_1$, and 
the sbottoms either remain stable or 
decay into other light hadrons (e.g. via $R$-parity violating
couplings) and go into the 
$b$-jets.  Gluino-pair production thus gives rise to inclusive $b$-quark
cross section.   The mass range of gluino is $m_{\tilde{g}}=12-16$ GeV 
and sbottom $m_{\tilde{b}_1}=2-5.5$ GeV.  Such masses are chosen so that
both the total cross section and the transverse momentum spectrum
of the $b$-quark are reproduced.
Before Berger et al.'s work, there have been some studies in the light sbottom
and/or light gluino scenario \cite{new}.  However, such a scenario cannot be
ruled out, unless there exists a sneutrino of at most 1--2 GeV.

Such a scenario easily contradicts other experiments, 
especially, the $Z^0$-pole data because of the light sbottom.
However, it can avoid the $Z$-pole constraints by tuning the coupling of 
$Z \tilde{b}_1 \tilde{b}_1^*$ to zero by choosing 
a specific mixing angle $\theta_b$ of $\tilde{b}_L$ and $\tilde{b}_R$:
$\sin^2\theta_b=\frac{2}{3} \sin^2\theta_W$, where $\theta_W$ is the Weinberg
mixing angle.
In spite of this, subsequent studies \cite{cao,cho,baek} showed that 
such light gluino and sbottom will still contribute significantly to 
$R_b$ via one-loop gluino-sbottom diagrams.  
In order to suppress such contributions, 
the second $\tilde{b}_2$ has to be lighter than about 180 GeV 
(at $5\sigma$ level) with the corresponding mixing angle in order 
to cancel the contribution of 
$\tilde{b}_1$ in the gluino-sbottom loop contributions to $R_b$.  
Although Berger et al.'s scenario is not ruled out, it certainly needs a lot
of fine tuning in the model.
In other words, instead of saying this scenario is fine-tuned, we 
can say that so far the light gluino and light sbottom scenario is not 
ruled out.  It definitely deserves more studies, no matter whether it was used
to explain the excess in hadronic bottom-quark production or not.

The light gluino and light sbottom scenario will possibly give rise to 
other interesting signatures, e.g., decay of $\chi_b$ into the light sbottom
\cite{lee}, enhancement of $t\bar t b\bar b$ production at hadron 
colliders \cite{rain}, decay of $\Upsilon$ into a pair of light sbottoms 
\cite{cla}, and affecting the Higgs decay \cite{hdecay}.
In a previous work \cite{yee}, 
we calculated the associated production of a gluino-pair
with a $q\bar q$ pair and compared to the standard model (SM)
 prediction of $q\bar q b\bar b$
at both LEPI and LEPII (here $q$ refers to the sum over $u,d,c,s,b$).
We found that at LEPII the $q\bar q \tilde{g} \tilde{g}$ production
cross section is about $40-20\,\%$ of the SM production of $q\bar q b\bar b$,
which may be large enough to produce an observable 
excess in $q\bar q b\bar b$ events \cite{yee}.  This is rather 
model-independent, independent of the mixing angle in the sbottom, and
is a QCD process.

In this work, we present another interesting channel in $Z$ decay 
in the light gluino and light sbottom scenario:
\begin{equation}
Z \to b \tilde{b}^*_1 \tilde{g}\; +\; \bar b \tilde{b}_1 \tilde{g}\; ; \qquad 
\mbox{followed by}\;\; 
\tilde{g} \to b \tilde{b}^*_1 \,/ \, \bar b \tilde{b}_1 \;.
\end{equation}
Since the gluino is a Majorana particle, so it can decay either
into $b \tilde{b}^*_1$ or $\bar b \tilde{b}_1$.  The final state can be 
$ bb \tilde{b}_1^* \tilde{b}_1^*$, $\bar b \bar b \tilde{b}_1 \tilde{b}_1$, 
or $ b \bar b \tilde{b}_1 \tilde{b}_1^*$.
This channel, unlike the one mentioned 
in the previous paragraph, depends on the mixing angle of $\tilde{b}_L$ and
$\tilde{b}_R$ in the $b \tilde{b}_1^* \tilde{g}$ coupling.  

The hadronic branching ratio of this channel will be shown to be $(3.4-2.5)
\times 10^{-3}$ for $\sin 2\theta_b >0$ and $(1.4-1.1) \times 10^{-3}$
for $\sin 2\theta_b <0$, and for 
$m_{\tilde{g}}=12-16$ GeV and $m_{\tilde{b}_1}=3$ GeV,
which is of order of the size of the uncertainty in $R_b$.  
The process is the supersymmetric analog of $Z\to b \bar b  g$, but 
kinematically they are very different because of the finite mass of the gluino 
and sbottom.
A typical event consists of an energetic prompt bottom-jet back-to-back with 
a ``fat'' bottom-jet, which consists of a bottom quark and two bottom squarks.
If such events cannot be distinguished from the prompt $b\bar b$ events,
they may increase the $R_b$ measurement ($R_b^{\rm exp}=0.21646\pm 0.00065$
\cite{lep-wg})
with a hadronic branching ratio of $(1-3)\times 10^{-3}$.
If the ``fat'' bottom jet can be distinguished from the ordinary bottom jet,
then this kind of events would be very interesting on their own.  It is
a verification of the light gluino and light sbottom scenario. 
Furthermore, if the flavor of the bottom quarks can be identified, the ratio
of $bb:\bar b \bar b:b\bar b$ events can be tested (theoretically it is
$1:1:2$) \cite{berger}.

The paper is organized as follows.  In the next section, we present the 
calculation, including the decay of the gluino into $b \tilde{b}_1^*$ or
$\bar b \tilde{b}_1$.  In Sec. III, we show the results and various 
distributions that verify the ``fat'' bottom jet.  
We conclude in Sec. IV.
There is an analog in hadronic collisions, $p \bar p \to b \tilde{b}_1^*
\tilde{g}$ followed by 
$\tilde{g} \to b \tilde{b}^*_1 \,/ \, \bar b \tilde{b}_1$.  Thus, it also
gives rise to two hadronic bottom jets.  However, in hadronic environment
it is very difficult to identify the ``fat'' bottom jet.  We believe it
only gives a small correction to the inclusive bottom cross section.

\section{Formalism}
The interaction Lagrangian among the bottom quark, sbottom, and gluino
is given by
\begin{equation}
\label{lag}
{\cal L} \supset \sqrt{2}g_s
[\tilde b_{1,i}^\dagger \bar{\tilde{g^a}} 
        (\sin\theta_bP_L+\cos\theta_bP_R)T_{ij}^a b_j 
    +\hbox{ h.c. }]  
\;,
\end{equation}
where the lighter sbottom $\tilde b_1$ is a superposition
$ \tilde b_1=\sin\theta_b \tilde b_L +\cos\theta_b \tilde b_R$ of the
left- and right-handed states via the mixing angle $\theta_b$.
As mentioned above, the vanishing of the  $Z  \tilde b_1 \tilde b_1^*$ 
coupling requires $ g_L \sin^2\theta_b +g_R \cos^2\theta_b =0$,
where 
$ g_L=-{1\over2}+{1\over3}\sin^2\theta_W$ and 
$ g_R={1\over3}\sin^2\theta_W$.
It implies $\sin^2\theta_b={2\over3}\sin^2\theta_W$.

\subsection{Primary Production}
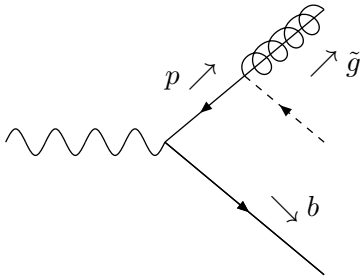
\begin{figure}[th!]
\begin{center}
\begin{picture}(100,100)(100,0)
\Photon(50,50)(110,50){5}{4}
\ArrowLine(110,50)(170,0)
\ArrowLine(110,50)(170,0)        \Text(150,25)[l]{$\searrow b$}
\Line(170,100)(140,75)           \Text(165,80)[l]{$\nearrow\tilde g$}
\ArrowLine(140,75)(110,50)       \Text(130,75)[r]{$p \nearrow $} 
\Gluon(140,75)(170,100){5}{4}     
\DashArrowLine(170,50)(140,75){3}  
\end{picture}
\caption{\small \label{fig1}
The Feynman diagram for the process
$Z\to b \tilde b_1^* \tilde g$.
}
\end{center}
\end{figure}

Even a perfect cancellation in the amplitude $Z\to \tilde b_1 \tilde
b_1^*$, the $Z$ boson can still decay at tree level 
into $b \tilde b_1^* \tilde g$ (or its conjugated channel) 
as shown in Fig.~\ref{fig1}.
The Feynman amplitude is 
\begin{equation}
{\cal  M}= \sqrt{2}\, g_s\, g_Z\, \bar u(b){\not \epsilon}_Z\,(g_LP_L+g_RP_R)
\,    {-\not p +m_b \over p^2 - m_b^2 } \,
  (\sin\theta_bP_R+\cos\theta_bP_L)\, T_{ij}^a \,v(\tilde g)  \ , 
\label{eq:decaya}
\end{equation}
where $P_{L,R}=( 1\mp \gamma^5)/2$, $g_Z=g_2/\cos\theta_W$, 
and  $i,j,a$ correspond  to the color indices 
of the final-state particles $b$, $\tilde {b}_1^*$ and $\tilde g$, 
respectively.
We can tabulate the complete formula of the transitional probability,
summing over the initial- and final-state spin polarizations or helicities,
and colors, as
\begin{equation}
\label{m2}
 \sum|{\cal M}|^2 ={16 g_s^2 g_Z^2   \over
(p^2-m_b^2)^2}
(N_0 + m_b m_{\tilde g} N_1\sin2\theta_b  + m_b^2 N_2 )  
\ ,\end{equation}
\begin{equation}
 \begin{array}{l}
N_0=(g_L^2 \sin^2\theta_b +g_R^2\cos^2\theta_b)    
[4 \tilde g\cdot p \; b\cdot p +p^2 \,\tilde g\cdot b\;(p^2-m_b^2-2s)/s
+2\tilde g\cdot p \; p\cdot Z m_b^2/s
]  \\
N_1=3(p^2+m_b^2) g_L g_R +(g_L^2+g_R^2)(p\cdot b +2 p\cdot Z \; b\cdot Z/s)  \\
N_2= 6 g_Lg_R \,\tilde g\cdot p + (g_R^2 \sin^2\theta_b +g_L^2\cos^2\theta_b)
        (\tilde g \cdot b + 2\tilde g\cdot Z \; b \cdot Z/s)
           \ , \end{array} \end{equation}
where $s=M_Z^2$.
Here the momenta of the particles are denoted by their corresponding
symbols. 
We use $p$ to denote 
the momentum of the virtual $\bar b$, which  turns into $\tilde g$ and 
$\tilde b_1^*$ (i.e. $p=\tilde{g} + \tilde{b}_1^*$). 

One can integrate the exact 3-body phase space to find the decay rate,
\begin{equation}
 d\Gamma(Z\to b \tilde b_1^* \tilde{g} ) 
={1\over3} \sum|M|^2  
 {\sqrt{s}\over \pi^3} {dx_b d x_{\tilde b}\over 256}\ . \end{equation}
The scaling variables of the 3-body phase space are defined by
\begin{equation}
           x_b=2 E_b/M_Z\ , \ 
  x_{\tilde b}=2 E_{\tilde{b}_1^*}/M_Z \ , \ \hbox{ and }\ 
  x_{\tilde g}=2 E_{\tilde g}/M_Z \ , \end{equation}
with the energies $E_i$ measured in the $Z$ rest frame, and 
$x_b + x_{\tilde b} + x_{\tilde g}=2$.
The ratios of the mass-squared are 
\begin{equation} \mu_b =m_b^2/M_Z^2 \ ,\ 
   \mu_{\tilde b} =m_{\tilde{b}_1}^2/M_Z^2 \ ,\ \hbox{ and }
   \mu_{\tilde g} =m_{\tilde g}^2/M_Z^2 \ .\end{equation}
The region of the phase space is limited by
\begin{equation}
  2\sqrt{\mu_b} \le x_b \le 1+\mu_b-\mu_{\tilde b}-\mu_{\tilde g}
                                -2\sqrt{\mu_{\tilde b}\mu_{\tilde g}} \ ,
\end{equation}
$$   x_{\tilde b} {\ }^{<}_{>} \hbox{$1\over2$} (1-x_b+\mu_b)^{-1}[
   (2-x_b)(1+\mu_b+\mu_{\tilde b}-\mu_{\tilde g}-x_b)
\pm(x_b^2-4\mu_b)^{1\over2}
         \lambda^{1\over2}(1+\mu_b-x_b,\mu_{\tilde b},\mu_{\tilde g})] $$
with the function $\lambda(a,b,c)=a^2+b^2+c^2-2ab-2bc-2ca$.
The scalar dot-products can be expressed in terms of the scaling variables as
$$ p^2=s(1+\mu_b-x_b) \ ,\quad
   \tilde g \cdot b=\hbox{$1\over2$}s
  (1-x_{\tilde b}+\mu_{\tilde b}-\mu_{\tilde g}-\mu_b)  $$
$$ b\cdot p=\hbox{$1\over2$} s(x_b-2\mu_b) \ ,\quad 
   \tilde g\cdot p = \hbox{$1\over2$} s(1-x_b
           -\mu_{\tilde b}+\mu_{\tilde g}+\mu_b) $$

The calculation for the charge-conjugated process 
$Z\to \bar b \tilde b_1 \tilde g$
can be repeated in a straightforward manner. 
Eqs. (\ref{eq:decaya}) and (\ref{m2}) remain valid if we make 
the substitutions 
$b      \leftrightarrow \bar b$,
$\tilde b_1^* \leftrightarrow  \tilde b_1$.

\subsection{Decay of gluino}

Since the gluino so produced will decay promptly into $b \tilde{b}_1^*$ or
$\bar b \tilde{b}_1$, the event ends up with the final states
$b      b\tilde b_1^* \tilde{b}_1^*$, 
$b \bar b\tilde b_1   \tilde{b}_1^*$, or
$\bar b \bar b \tilde b_1   \tilde{b}_1$.
In the minimal hypothesis that the sbottom hadronizes completely in the
detector, it behaves like a hadronic jet. The final configuration
includes $bb+ 2j$, $b\bar b+2j$, and $\bar b \bar b +2j$ at the parton 
level.  We 
will show below that the $2j$ most of the time goes together with the 
softer $b$, and therefore makes the $b$ look ``fat''.
The complete jet structure requires the full helicity calculation 
following the decay chain $Z\to b\tilde b_1^*\tilde g$ and
$\tilde g \to b \tilde{b}_1^*$ or
$\bar b \tilde{b}_1$.
Based on Feynman rules for the Majorana fermions, we replace 
$v(\tilde g)$ in the above Eq.~(\ref{eq:decaya}) by
\begin{equation}
\label{ch1}
{\rm Ch. 1:}\;\;\;\;\;  v(\tilde{g}) \longrightarrow \; -
\sqrt{2} g_s  \, T^a \, 
\frac{ -\not \tilde{g} + m_{\tilde{g}} }
     { \tilde{g}^2 - m^2_{\tilde{g}} +i\Gamma_{\tilde g} m_{\tilde g}}
\left( \sin \theta_b P_L + \cos\theta_b P_R \right ) v(\bar b) \ ,
\end{equation}
for the process $\tilde g \to \bar b \tilde b_1$.
Similarly, we replace $v(\tilde g)$ in  Eq.~(\ref{eq:decaya}) by
\begin{equation}
\label{ch2}
{\rm Ch. 2:}\;\;\;\;\;   v(\tilde{g})  \longrightarrow 
\sqrt{2} g_s  \, T^a \,
\frac{ -\not \tilde{g} + m_{\tilde{g}} }
     { \tilde{g}^2 -m^2_{\tilde{g}} +i\Gamma_{\tilde g} m_{\tilde g}}
\left( \sin \theta_b P_R + \cos\theta_b P_L \right ) v(b) \ ,
\end{equation}
for the process $\tilde g \to  b \tilde b_1^*$.
We use the narrow-width approximation to calculate the on-shell
gluino propagator
\begin{equation}
\frac{1}{ (\tilde{g}^2 - m^2_{\tilde{g}})^2 + \Gamma^2_{\tilde{g}}
m^2_{\tilde{g}} } 
  \longrightarrow \frac{\pi}{m_{\tilde{g}} \Gamma_{\tilde{g}} } \,
\delta( \tilde{g}^2 - m^2_{\tilde{g}} ) \;,
\end{equation}
where $\tilde{g} = b + \tilde{b}_1^*$ or $\bar b +\tilde{b}_1$.
Assuming the gluino only decays into $b \tilde{b}_1^*$ and 
$\bar b \tilde{b}_1$, we find the decay width of the gluino is
\begin{equation}
\Gamma_{\tilde{g}} = \hbox{$1\over4$} (\alpha_s/m_{\tilde g})\,
\lambda^{1\over2}(1, m^2_b/m^2_{\tilde g}, m^2_{\tilde{b}_1}/m^2_{\tilde g})\,
(m^2_{\tilde{g}} + m^2_b - m^2_{\tilde{b}_1} 
  + 2 m_{\tilde{g}}m_{b}  \sin 2\theta_b)
\;.
\end{equation}

Since we have already assumed CP invariance in Eq. (\ref{lag}), 
the event distributions of a pair of CP-conjugated variables 
are the same.

\section{Results}

We first list the input parameters in our study
\[
m_b=4.5\,{\rm GeV}\,, \;\; m_{\tilde{b}_1}=3\,{\rm GeV}\,, \;\;
\sin\theta_b = \sqrt{\frac{2}{3} \sin^2\theta_W}\,, \;\;
\cos\theta_b = \pm \sqrt{ 1- \frac{2}{3} \sin^2\theta_W}\,.
\]
The scale $Q$ that we used in the running strong coupling constant is
evaluated at $\alpha_s(Q = M_Z/2)$.
\footnote{
The difference in $\alpha_s$ between including and not including the light 
gluino and sbottom in the running of $\alpha_s$ from $Q=M_Z$ to $M_Z/2$ 
is only 3\%.  Thus, we neglect the effect of light gluino and sbottom 
in the running of $\alpha_s$.  Refs. \cite{berger,cla} also estimated
the effect of including the light gluino in the running of $\alpha_s$ in their
studies.
A recent work \cite{chiang} studied the running of $\alpha_s$ from low-energy
scales such as $m_\tau$ to $M_Z$ including a light gluino and a light sbottom.
However, it cannot rule out the existence of such light particles from
current data.  
}

\begin{figure}[th!]
\includegraphics[width=6in]{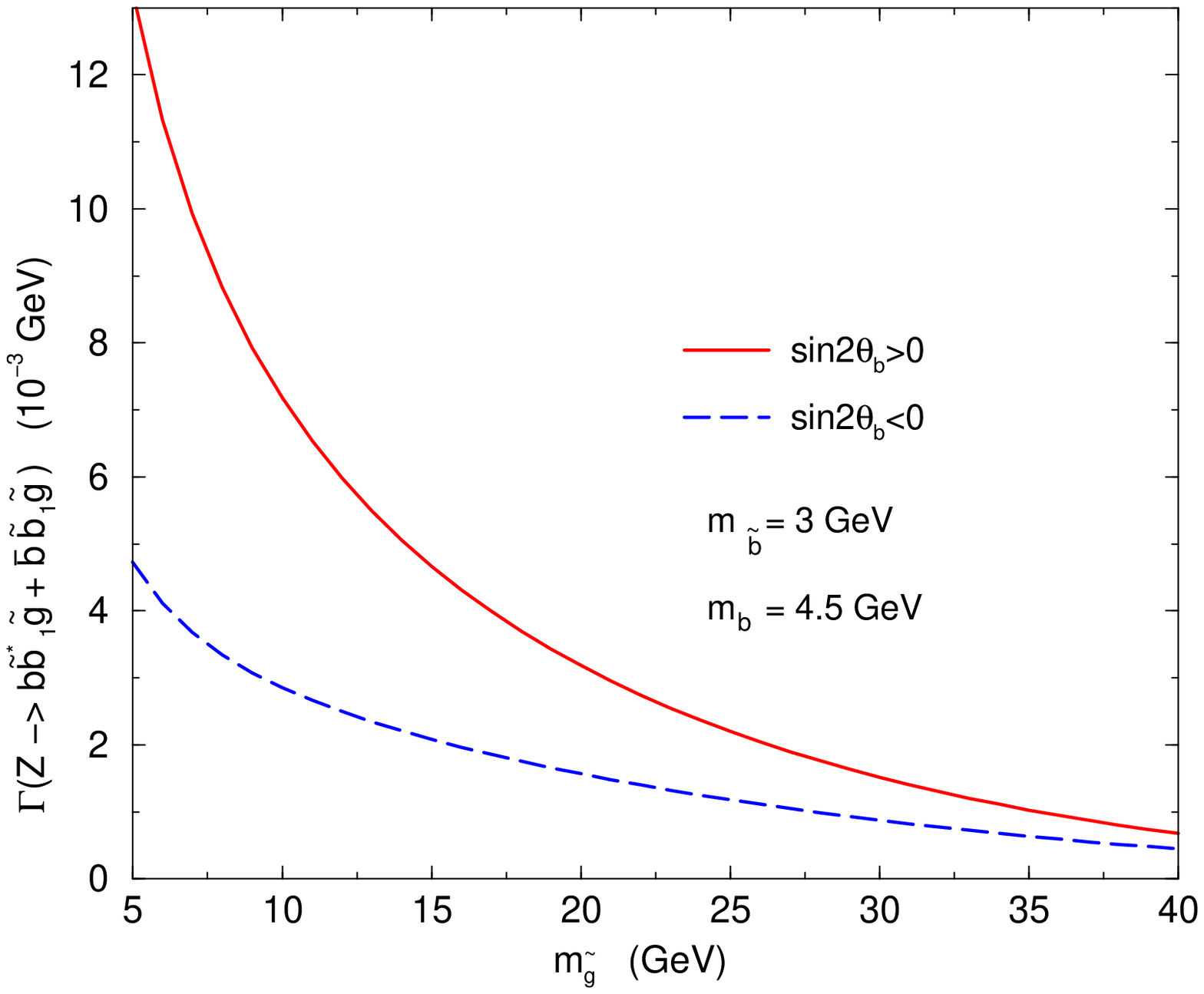}
\caption{ \small
\label{pw}
Partial width of $Z\to b \tilde b_1^* \tilde g \,+ \,
\bar b \tilde{b}_1 \tilde{g}$
versus $m_{\tilde g}$ for $m_{\tilde b_1}=3$ GeV and
$m_b=4.5$ GeV.
}
\end{figure}

We show in Fig. \ref{pw} the partial width of the channel 
$Z \to b \tilde b^*_1 \tilde g \,+\, \bar b \tilde{b}_1 \tilde{g}$ 
versus the gluino mass $m_{\tilde g}$ 
for two different sign choices $\sin2\theta_b {\ }^>_<0$. 
Numerically, the effect of $m_b$ is not negligible at $\sqrt{s}=M_Z$. 
Given that the total hadronic width of the $Z$ boson is 1.745 GeV
\cite{lep-wg}, the hadronic branching fraction of the
process $Z \to b \tilde b^*_1 \tilde g \,+\,
\bar b \tilde{b}_1 \tilde{g}$ is $(3.4 - 2.5) \times 10^{-3}$
for $\sin 2\theta_b >0$ and $(1.4 - 1.1) \times 10^{-3}$ for 
$\sin 2\theta_b <0$, and $m_{\tilde{g}}=12-16$ GeV.
Thus, this hadronic branching ratio is at the level of, or even larger than,
the uncertainty in the $R_b$ measurement ($R_b^{\rm exp}=0.21646\pm 0.00065$).
If it cannot be distinguished from the prompt $b\bar b$ events, it will affect
the precision measurement on the $b\bar b$ yield at LEP I.

\begin{figure}[th!]
\includegraphics[width=4in]{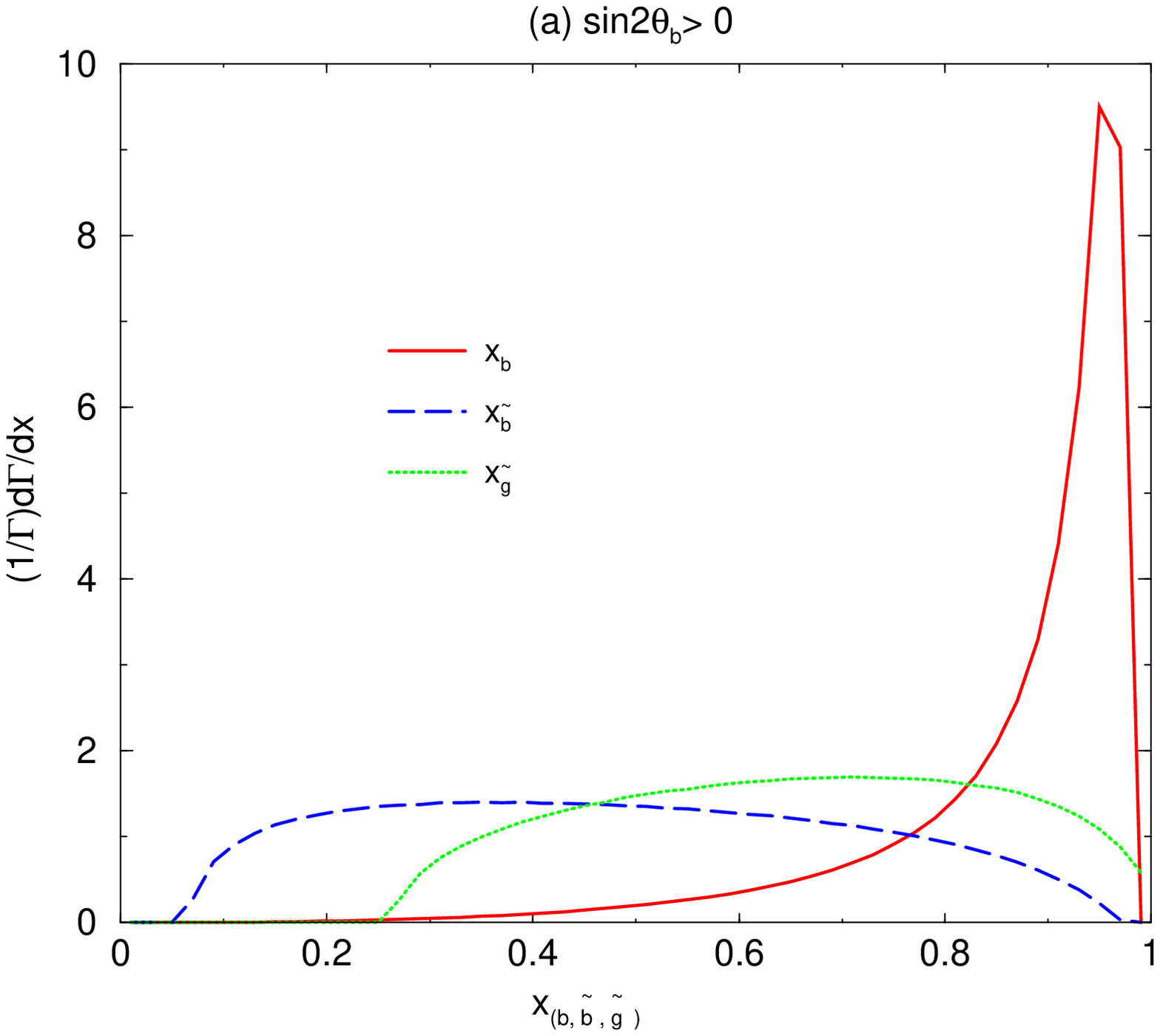}

\vspace{0.6in}

\includegraphics[width=4in]{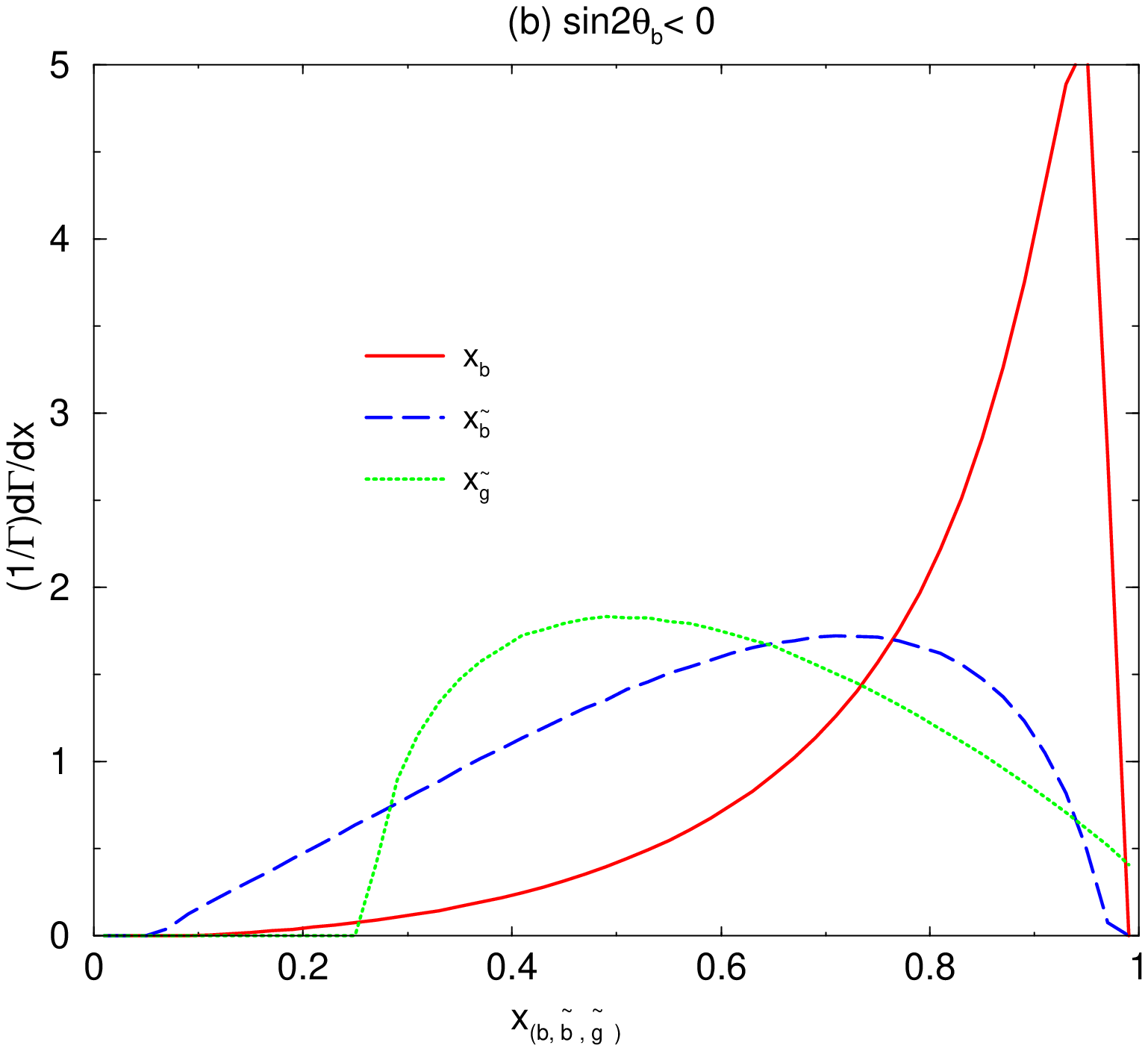}
\caption{ \small
\label{xb}
Normalized energy spectra of the $b$ ($x_b$), sbottom ($x_{\tilde b}$), 
and gluino ($x_{\tilde g}$) in the decay $Z \to b \tilde{b}_1^* \tilde{g}$.
(a) $\sin 2 \theta_b > 0$ and (b) $\sin 2 \theta_b <0$.}
\end{figure}

In the following, we study the event topology to examine the difference
from the prompt $b\bar b$ production, which essentially consists of 
two back-to-back clean bottom jets with energy equal to $M_Z/2$.
In Fig. \ref{xb}, we show the energy distributions, in terms of 
dimensionless variables $x_b$, $x_{\tilde b}$, $x_{\tilde{g}}$, of
the prompt $b$, sbottom, and gluino, respectively.
The prompt $b$ has a fast and sharp energy distribution as expected, 
but the gluino and the sbottom have slower and flatter
energy spectra.  We also note that the spectra are different between
$\sin 2 \theta_b >0$ and $<0$.
These features are very different from the prompt
$b\bar b$ production including QCD correction, in which both $b$ and $\bar b$
are very energetic and the gluon is quite soft.  

\begin{figure}[th!]
\includegraphics[width=4in]{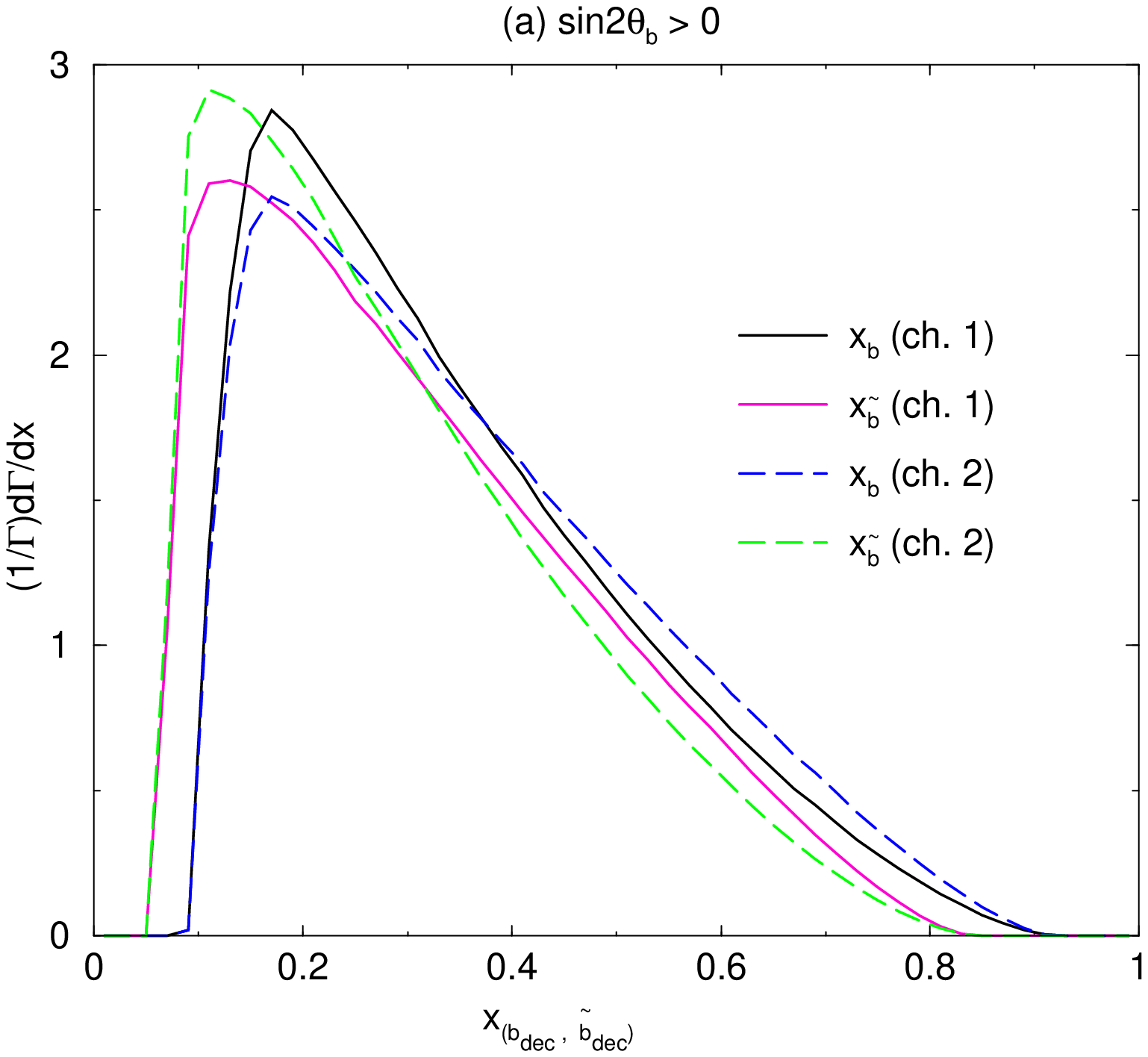}

\vspace{0.6in}

\includegraphics[width=4in]{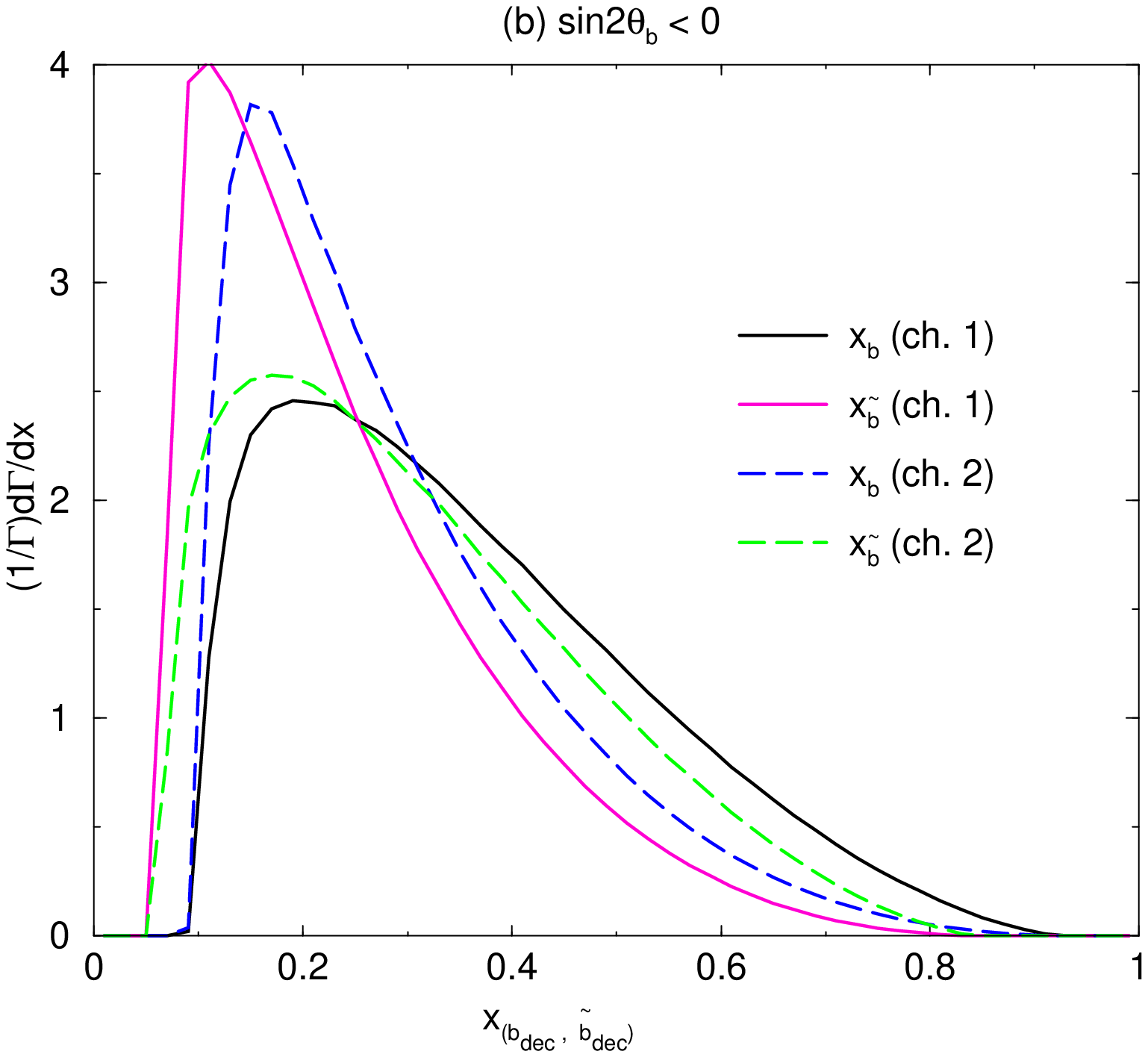}
\caption{ \small
\label{xbdec}
Normalized energy spectra $x_{b_{\rm dec}}$ and
$x_{\tilde{b}_{\rm dec}}$, in which the bottom and sbottom are the subsequent
decay products of the gluino.
(a) $\sin 2 \theta_b > 0$ and (b) $\sin 2 \theta_b <0$.
Here ``ch. 1'' and ``ch. 2'' refer to the decay channels of the gluino
in Eqs. (\ref{ch1}) and (\ref{ch2}), respectively.
}
\end{figure}

In Fig. \ref{xbdec}, we show the energy spectra for the decay products,
$b_{\rm dec}$ and $ \tilde{b}_{\rm dec}$, of the gluino.  
Since gluino is a Majorana
particle, it decays into either $b \tilde{b}_1^*$ or $\bar b \tilde{b}_1$.
Although there are some differences between these two decay modes because of
the difference in the coupling, in both modes the $b_{\rm dec}$ and 
$\tilde{b}_{\rm dec}$ are rather soft.
We also note that the spectra are different between $\sin 2 \theta_b >0$ and
$<0$.
Therefore, just by looking at the prompt $b$ and the secondary $b_{\rm dec}$
the energy spectra are very different from the prompt $b\bar b$ production.
However, if the first and the second sbottoms go very close with the
secondary $b_{\rm dec}$ and cannot be separated experimentally, and the
sbottoms deposit all their energies in the detector, then the event will
mimic the prompt $b\bar b$ event.
Thus, it is important to look at the angular separation among the final-state
particles.

\begin{figure}[th!]
\includegraphics[width=6in]{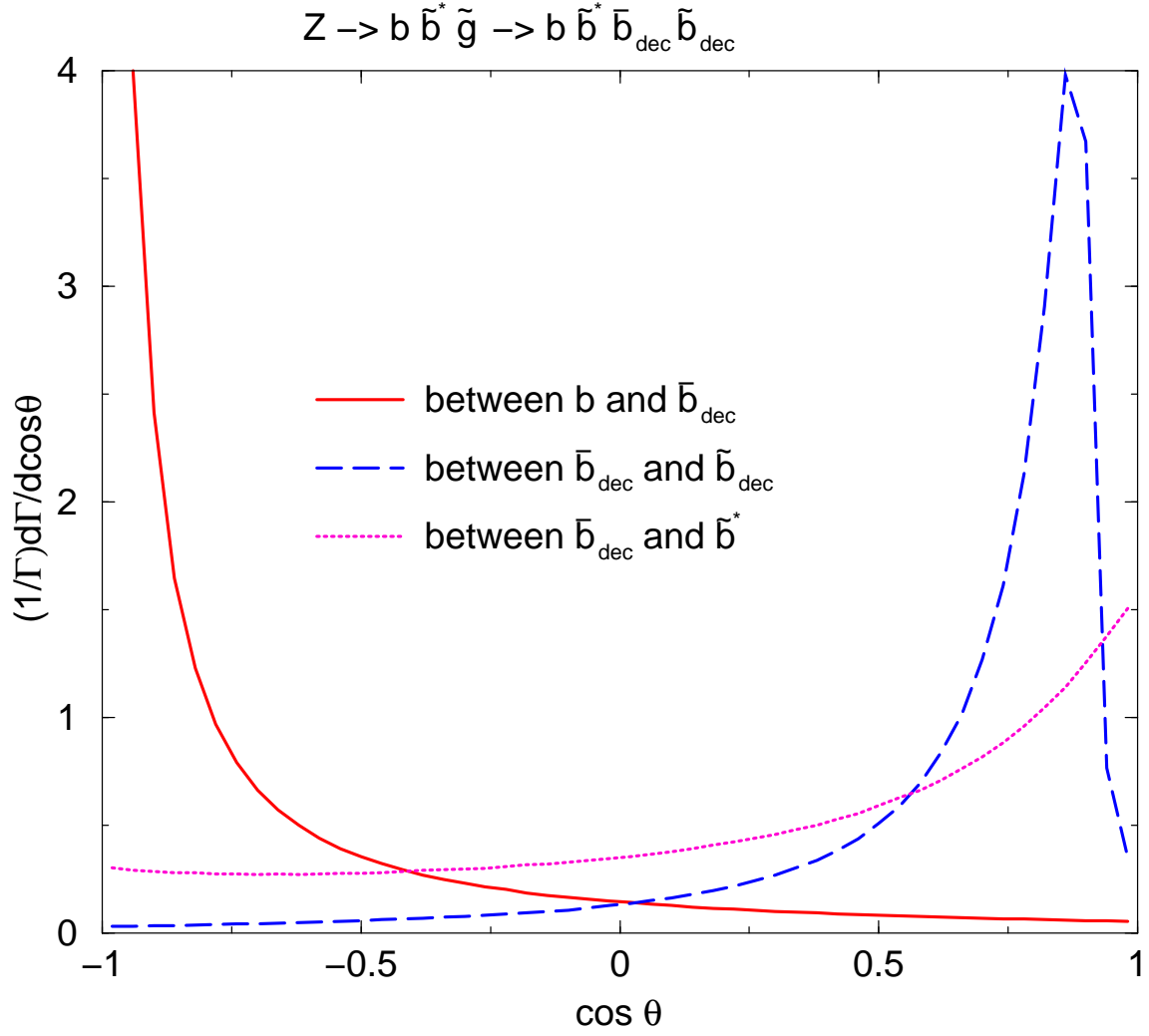}
\caption{ \small
\label{cos}
Normalized spectra of the cosine of the angles between various pairs
of final-state partons in the decay process $Z\to b\tilde b_1^*\tilde g$,
followed by $\tilde g \to \bar b \tilde{b}_1$.
Here $\bar b_{\rm dec}$ and $\tilde{b}_{\rm dec}$ denote the decay products
of the gluino.
}
\end{figure}

We show the cosine of the angles between the primary $b$ and
the $\bar b_{\rm dec}$, between $\bar b_{\rm dec}$ and $\tilde{b}_{\rm dec}$,
and between $\bar b_{\rm dec}$ and $\tilde{b}_1^*$ in Fig. \ref{cos}.
Here we only show the spectra for the case $\sin 2 \theta_b>0$ and gluino
decay Ch. 1, because for $\sin 2 \theta_b>0$ or $<0$, Ch. 1 or Ch. 2, the 
spectra are very similar.
We can immediately see that the primary $b$ is back-to-back with the
secondary $\bar b_{\rm dec}$ from gluino decay.  The $\bar b_{\rm dec}$ and 
$\tilde{b}_{\rm dec}$ are very much close to each other that the cosine
of the angle between them is peaked at $0.8-0.9$.  
The cosine of the angle between $\bar b_{\rm dec}$ and $\tilde{b}_{1}^*$ has
a broader distribution, but still peaks in the $\cos\theta=1$ region.  Thus,
we have the following picture.  The decay products, 
$\bar b_{\rm dec}$ and $\tilde{b}_{\rm dec}$, and the primary $\tilde{b}_1^*$ 
combine to form a wide or ``fat'' bottom-like jet.  This ``fat'' bottom
jet is back-to-back to the primary bottom jet, which has an energy close to
$M_Z/2$.

Here we comment on the possibility that the channel that we consider here
may affect the $R_b$ measurement, based on two criteria.  
First, one of the
bottom jet in the channel under  consideration is ``fat''.  
If the two sbottoms cannot be 
separated from the bottom, the resulting bottom jet will just look like
a fat bottom jet and may affect $R_b$.
Second, whether the energy in this fat bottom jet equals to half of the 
$Z$ mass or not.
As mentioned by Berger et al. \cite{berger}, the sbottom can either decay
into light hadrons or escape unnoticed from the detector.  If the sbottoms
escape the detection (which means that they do not deposit enough kinetic
energy in the detector material for detection), the fat bottom jet would
have an energy much less than $M_Z/2$.  The final-state would be 
two bottom jets (one energetic and one much less energetic) 
plus missing energy, and thus would not affect $R_b$.
Neverthesless, this is a very interesting signal on its own.
On the other hand,  if the sbottoms deposit all their kinetic energy in the
detector, the measured energy of the fat bottom jet would be close to
$M_Z/2$.  In this case, it may affect the measurement of $R_b$. In fact, it
would increase $R_b$. But if the fat bottom jet could be distinguished from 
the normal bottom jet, the present channel is also interesting on its own.
According to a study on the light gluino \cite{jack}, a sbottom of mass 
$2-5.5$ GeV, if similar to gluino, will likely deposit most of its
kinetic energy in the detector.  If this is the case the signal would be
two back-to-back bottom jets, one of which is ``fat'' or wide, with no
or little missing energy.

\section{Conclusions}
We show that the light-sbottom-gluino scenario predicts the production
of $b      b\tilde b_1^* \tilde{b}_1^*$,
$b \bar b\tilde b_1   \tilde{b}_1^*$, and 
$\bar b \bar b\tilde b_1   \tilde{b}_1$
at the $Z$ pole, with a
branching fraction of order of $10^{-3}$, depending on the gluino mass and 
the sign of the mixing angle.
The event topology is very different from the prompt $b\bar b$ production. 
Depending on whether the sbottoms deposit little or almost all of their
energies in the detector, the signal would be
very different.  If the sbottoms
escape the detector unnoticed, the final-state would be 
two bottom jets (one energetic and one much less energetic) 
plus missing energy.
On the other hand,  if the sbottoms deposit all their kinetic energy in the
detector,  the final state will be two bottom jets, one of which is fat.
In this case, it may increase the measurement of $R_b$.
But if the fat bottom jet could be separated from 
the normal bottom jet, it is a distinct signal.
These two kinds of signals may well be
hidden in the LEP I data, waiting for deliberate search.

One special feature of the Majorana nature of the gluino predicts
a ratio of 1:1:2 for the rates of $bb\,:\, \bar b \bar b\,:\,b\bar b$
\cite{berger}. 
However, one needs to look
for the charged  modes $B^+B^+$ or $B^-B^-$   to avoid effects due to
$B^0$-$\bar B^0$ oscillation.

\section*{Acknowledgment}
This research was supported in part by the National Center for Theoretical
Science under a grant from the National Science Council of Taiwan
R.O.C., and in part by US DOE (Grant number DE-FG02-84ER40173).


\end{document}